\documentclass{LMCS}

\usepackage{enumerate}
\usepackage{hyperref}
\usepackage{amssymb}
\usepackage{latexsym}

\newcommand{\N}{{\mathbb N}}

\newcommand{\TN}{2^{\N}}

\newcommand{\A}{{\mathcal A}}
\newcommand{\B}{{\mathcal B}}
\newcommand{\T}{{\mathcal T}}
\newcommand{\C}{{\mathcal C}}

\newcommand{\ob}{\hat{b}}
\newcommand{\fr}{{^\frown}}
\newcommand{\la}{\langle}
\newcommand{\ra}{\rangle}

\newcommand{\ml}{Martin-L\"{o}f\ }

\newcommand{\Iff}{\ \iff\ }

\theoremstyle{plain}

\def\doi{6 (3:16) 2011}
\lmcsheading%
{\doi}
{1--16}
{}
{}
{Dec.~27, 2010}
{Sep.~23, 2011}
{}

\begin{document}

\title[Randomness and Capacity]{Algorithmic Randomness and Capacity of
  Closed Sets\rsuper*}

\author[P.~Brodhead]{Paul Brodhead\rsuper a}
\address{{\lsuper a}Indian River State College, Fort Pierce, Florida}
\email{brodhe@gmail.com}

\author[D.~Cenzer]{Douglas Cenzer\rsuper b}
\address{{\lsuper b}Department of Mathematics, University of Florida,
Gainesville, Florida 32611-8105}
\email{cenzer@ufl.edu}
\thanks{{\lsuper b}This research was partially supported by NSF
grants DMS 0532644 and 0554841 and 652372}

\author[F.~Toska]{Ferit Toska\rsuper c}
\address{{\lsuper{c,d}}University of Florida}
\email{toskaf@ufl.edu and swyman@ufl.edu>}

\author[Wyman]{Sebastian Wyman\rsuper d}


\titlecomment{{\lsuper*}A preliminary version of this paper appeared in the CCA
  2010 proceedings EPCTS 24}

\begin{abstract} 
We investigate the connection between measure, capacity and
algorithmic randomness for the space of closed sets.  For any
computable measure $m$, a computable capacity $T$ may be defined by
letting $T(Q)$ be the measure of the family of closed sets $K$ which
have nonempty intersection with $Q$. We prove an effective version of
Choquet's capacity theorem by showing that every computable capacity
may be obtained from a computable measure in this way. We establish
conditions on the measure $m$ that characterize when the capacity of
an $m$-random closed set equals zero. This includes new results in
classical probability theory as well as results for algorithmic
randomness. For certain computable measures, we construct effectively
closed sets with positive capacity and with Lebesgue measure zero.  We
show that for computable measures, a real $q$ is upper semi-computable
if and only if there is an effectively closed set with capacity $q$.
\end{abstract}

\maketitle


\section*{Introduction}The study of algorithmic randomness has been an
active area of research in recent years.  The basic problem is to
quantify the randomness of a single real number. Here we think of a
real $r \in [0,1]$ as an infinite sequence of 0's and 1's, i.e.\ as an
element in $\TN$. There are three basic approaches to algorithmic
randomness: the measure-theoretic approach of \ml tests,
the incompressibility approach of Kolmogorov complexity, and the
betting approach in terms of martingales.  All three approaches have
been shown to yield the same notion of (algorithmic) randomness. The
present paper will consider only the measure-theoretic approach.  A
real $x$ is \ml random if for any effective sequence $S_1,
S_2, \dots$ of c.~e. open sets with $\mu(S_n) \leq 2^{-n}$, $x \notin
\bigcap_n S_n$.  For background and history of algorithmic randomness we
refer to \cite{DH:book,Nies:book}.

The study of random sets and in particular of random closed sets is a
vibrant area in probability and statistics, with many applications in
science and engineering.  The notion of capacity plays an important
role here as a part of the analysis of imprecise or uncertain
observations, for example in intelligent systems. For background on
the theory of random sets see \cite{Ng06}.

In a series of recent papers \cite{BCDW07,BCRW08}, G. Barmpalias,
P. Brodhead, D. Cenzer, S. Dashti, J.B. Remmel and R. Weber have
defined a notion of algorithmic randomness for closed sets and
continuous functions on $2^{\N}$. Here the Polish space $2^{\N}$ is
equipped with usual product topology and has a basis of clopen sets.
Definitions are given below in section \ref{sec1}. The space $\C$ of
closed subsets of $2^{\N}$ has the \emph{hit-or-miss} or \emph{Fell}
topology which is also described in section \ref{sec1}. In general
when we discuss closed sets in this paper we are refering to closed
subsets of $2^{\N}$. 

The study of randomness for closed sets and continuous functions has
several interesting aspects concerning properties of those sets and
properties of the members of such sets.  The topological and
measure-theoretic properties of effectively random closed sets has
been studied. For example, it is shown in \cite{BCDW07} that every
effectively random closed set is perfect and has Lebesgue measure 0.
The complexity of effectively random closed sets as subsets of
$2^{\N}$ was considered in \cite{BCDW07}, where it was shown that no
effectively closed ($\Pi^0_1$) set is random but there is a random
$\Delta^0_2$ closed set.

The members of a closed set are reals and hence we can study the
complexity of the members of an effectively random closed set. The
following results were obtained in \cite{BCDW07}. Every effectively
random closed set contains a random member but not every member is
random. Every random real belongs to some random closed set. Every
effectively random $\Delta^0_2$ closed set contains a random
$\Delta^0_2$ member. Effectively random closed set contain no
computable elements (in fact, no $n$-c.~e.\ elements). It was shown in \cite{BCRW08}
that the set of zeroes of an effectively random continuous function is
an effectively random closed set. 

Just as an effectively closed set in $\TN$ may be viewed as the set of
infinite paths through a computable tree $T \subseteq \{0,1\}^*$, an
algorithmically random closed set in $\TN$ may be viewed as the set of
infinite paths through an algorithmically random tree
$T$. Diamondstone and Kjos-Hanssen \cite{DK09,KH09} give an
alternative definition of algorithmic randomness for closed sets
according to the Galton-Watson distribution and show that this
definition produces the same family of algorithmically random closed
sets. 

We note that in probability theory a random closed subset of a
topological space $X$ is considered a random variable which takes on
the values in the space $\C(X)$ of closed subsets of $X$. That is, let
$(\Omega,\A,P)$ be a \emph{probability space} with underlying
topological space $\Omega$, $\sigma$-algebra $\A \subseteq {\mathcal
  P}(X)$ and measure $P$ such that $P(\Omega) = 1$ and $P(S)$ is
defined for all sets $S \in \A$.  For example, we might have $\Omega =
2^{\N}$, $\A$ the family of Borel subsets of $2^{\N}$, and $P$ the
standard Lebesgue measure.  The map $X$ induces a probability measure
$P_X$ on $\C(X)$ given by $P(X^{-1}(S))$. Classically, the statement
that a random closed set has no computable elements means that the
collection of closed sets with no computable elements has measure one.
In effective randomness, there is a particular collection $R$ of
\emph{algorithmically random} closed sets which has measure one. In
this context, the statement that effectively random closed sets have
no computable elements is to say that the closed sets in $R$ have no
computable elements. The latter result of course implies the former,
but is stronger.

A random closed set is a specific type of random recursive
construction, as studied by Graf, Mauldin and Williams
\cite{GMW88}. McLinden and Mauldin \cite{MM09} showed that the
Hausdorff dimension of a random closed set is $log_2(4/3)$, that is,
almost every closed subset of $2^{\N}$ has Hausdorff dimension
$log_2(4/3)$. It was shown in \cite{BCDW07} that every effectively
random closed set has box dimension $log_2(4/3)$.  The
\emph{effective} Hausdorff dimension of members of effectively random
closed sets is studied in \cite{DK09}. It is shown that every member
of an effectively random closed set has effective Hausdorff dimension
$\geq log_2(3/2)$ and that any real with effective Hausdorff dimension
$> log_2(3/2)$ is a member of some effectively random closed set.

In the present paper we will examine the notion of computable capacity
and its relation to computable measures on the space $\C$ of nonempty
closed sets. Given a domain $U$, a \emph{capacity} $\T$ is a
real-valued function defined on some $\sigma$-field of subsets of $U$,
which is closely related to measure.  $\T$ may be thought of as a
\emph{belief} function in the context of reasoning with uncertainty.
(See \cite[p.\ 71]{Ng06} and also \cite{Sh76}. ) The capacity $\T(A)$
for a set $A$ is the probability that a randomly chosen set $S$ is a
subset of $A$.

Choquet \cite{Ch55} developed the Choquet capacity for the space $\C$
of closed subsets of an infinite set $X$. A probability measure
$\mu^*$ on $\C$ induces a capacity $\T$ on $\C$ by defining the
capacity $\T(C)$ of a closed set $C$ to be $\mu^*(\{K \in \C: K
\subseteq C\})$.  Choquet's capacity theorem states that every
capacity $\T$ on $\C$ arises in this way from some measure $\mu^*$.

In section one, we give some basic definitions including the
definition of the space of $\C(X)$ of closed subsets of a computable
Polish space $X$. We present a family of computable measures on $\C$
which will lead to different notions of effective randomness for
closed sets.

In section two, we define the notion of computable capacity and show
how a measure on the space of closed sets induces a capacity.  An 
effective version of Choquet's capacity theorem is proved. 

The main theorem of section three gives conditions under which the
capacity $\T(Q)$ of a $\mu^*$-random closed set $Q$ is either equal to
$0$ or $> 0$. In particular, suppose that the measure $\mu_b$ on
$\{0,1,2\}^{\N}$ is defined so that, for all $\sigma \in \{0,1,2\}^*$,
$\mu_b(I(\sigma \fr i)) = b \cdot \mu_b(I(\sigma))$ for $i=0,1$ and define
the corresponding probability measure $\mu^*_b$ and capacity $\T_b$ on
the space $\C$ of closed sets and the corresponding capacity $\T_b$.
This means that for any node $\sigma$ in the tree $T_Q$, $\sigma$ has
unique extension $\sigma \fr 0$ in $T_Q$ with probability $b$, and
similarly $\sigma$ has unique extension $\sigma \fr 1$ with
probability $b$. Then we show the following. If $b \geq 1 -
\frac{\sqrt 2}2$, then every effectively $\mu^*_b$-random closed set
$Q$ has capacity $\T_b(Q) = 0$. It is important to note that, since
the random closed sets have measure one in the space $\C$ of closed
sets, this result implies that almost all closed sets have capacity
zero.  This is a \emph{new} result about the classical measure and capacity of
closed sets in general and not only about algorithmic randomness or
computability.

On the other hand, if $b < 1 - \frac{\sqrt 2}2$, then every
effectively $\mu^*_b$-random closed set $Q$ has capacity $\T_b(Q) >
0$, and hence almost every closed set has positive capacity.  A more
general result is given. 

In section four, we consider the capacity of effectively closed
sets. Fix computable reals $b_0$ and $b_1$ such that $0 < b_1 \leq
b_0$ and $b_0 + b_1 < 1$ and define the measure $\mu$ on
$\{0,1,2\}^{\N}$ so that for any $\sigma \in\{0,1,2\}^*$ and for $i
\in \{0,1\}$, $\mu(I(\sigma \fr i)) = b_i \cdot \mu(I(\sigma))$. Let
$\mu^*$ be the corresponding measure on $\C$ and let $\T$ be the
corresponding capacity. It is easy to see that for any effectively
closed set $Q$, $\T(Q)$ is an upper-semi-computable real. Conversely,
for any upper-semi-computable real $q$, there exists an effectively
closed set $Q$ with capacity $T(Q) = q$. We also show that if $b_0 =
b_1$, there exists an effectively closed set $Q$ with Lebesgue measure
zero and with positive capacity.

A preliminary version \cite{BC10} of this paper appeared in the
electronic proceedings of the conference CCA 2010. The current paper
contains several improvements and new results, including Theorems
\ref{th4a}, \ref{th4b}, \ref{th4c} and \ref{th5}. We thank the 
referees for very helpful comments. 

\section{Computable Measures on the Space of Closed
  Sets} \label{sec1}

We present an effective version of
Choquet's theorem connecting measure and capacity.

In this section, we describe the hit-or-miss topology on the space
$\C$ of closed sets, we define certain probability measures $\mu_d$ on
the space $\{0,1,2\}^{\N}$ and the corresponding measures $\mu^*_d$ on
the homeomorphic space $\C$. These give rise to notions of algorithmic
randomness for closed sets. 

 Some definitions are needed.  For a finite
string $\sigma \in\{0,1\}^n$, let $|\sigma| = n$.  Let $\lambda$
denote the empty string so that $|\lambda| = 0$. For two strings
$\sigma,\tau$, say that $\sigma$ is an \emph{initial segment} of
$\tau$ and write $\sigma \sqsubseteq \tau$ if $|\sigma| \leq |\tau|$
and $\sigma(i) = \tau(i)$ for $i < |\sigma|$. For $x \in \TN$, $\sigma
\sqsubset x$ means that $\sigma(i) = x(i)$ for $i < |\sigma|$. Let
$\sigma^{\frown} \tau$ denote the concatenation of $\sigma$ and $\tau$
and let $\sigma^{\frown} i$ denote $\sigma^{\frown}(i)$ for
$i=0,1$. For $\sigma \in \{0,1\}^*$ and $x \in \TN$, $\sigma \fr x =
(\sigma(0),\dots,\sigma(|\sigma|-1),x(0),x(1),\dots)$. Let $x\lceil n
=(x(0),\dots,x(n-1))$. Two reals $x$ and $y$ may be coded together
into $z = x \oplus y$, where $z(2n) = x(n)$ and $z(2n+1) = y(n)$ for
all $n$.
For a finite string $\sigma$, let $I(\sigma)$ denote $\{x \in
\TN:\sigma \sqsubset x\}$. We shall call $I(\sigma)$ the
\emph{interval} determined by $\sigma$. Each such interval is a clopen
set and the clopen sets are just finite unions of intervals. We let
$\B$ denote the computable Boolean algebra of clopen sets. Note that this is a countable atomless
Boolean algebra. 

A set $T \subseteq \{0,1\}^*$ is a \emph{tree} if it is closed under
initial segments.  For an arbitrary tree $T \subseteq \{0,1\}^*$, let
$[T]$ denote the set of infinite paths through $T$. It is well-known
that $P \subseteq \TN$ is a closed set if and only if $P = [T]$ for
some tree $T$. $P$ is a $\Pi^0_1$ class, or an effectively closed set, if $P = [T]$
for some computable tree $T$.

A closed set $P$ may be identified with a tree $T_P\subseteq
\{0,1\}^*$ where $T_P = \{\sigma: P \cap I(\sigma)
\neq \emptyset\}$. Note that $T_P$ has no dead ends. That is, if
$\sigma\in T_P$, then either $\sigma^{\frown}0 \in T_P$ or
$\sigma^{\frown}1\in T_P$. The complexity of the closed set $P$ is
generally identified with that of $T_P$.  Thus $P$ is said to be a
$\Pi^0_2$ closed set if $T_P$ is $\Pi^0_2$; in this case $P = [T]$ for
some $\Delta^0_2$ tree $T$.  The complement of an effectively closed set is
sometimes called a c.~e. open set. We remark that if $P$ is an effectively closed set,
then $T_P$ is a $\Pi^0_1$  set, but it is not, in general,
computable. For any  $\sigma \in \{0,1\}^*$ and any $Q \subseteq \TN$,
we let $\sigma \fr Q$ denote $\{\sigma \fr x: x \in Q\}$. 
There is a natural effective enumeration $P_0, P_1, \dots$ of the
effectively closed sets and thus an enumeration of the c.~e. open sets. Thus
we can say that a sequence $S_0,S_1,\dots$ of c.~e. open sets is
\emph{effective} if there is a computable function, $f$, such that
$S_n = \TN - P_{f(n)}$ for all $n$. For a detailed development of
effectively closed sets, see \cite{CR99}.

It was observed in \cite{BCDW07} that there is a natural isomorphism
between the space $\C$ of nonempty closed subsets of $\{0,1\}^{\N}$
and the space $\{0,1,2\}^{\N}$ (with the product topology) 
defined as follows. Given a nonempty closed
$Q\subseteq \TN$, let $T = T_Q$ be the tree without dead ends such
that $Q =[T]$. Let $\sigma_0, \sigma_1, \ldots$ enumerate the elements
of $T$ in order, first by length and then lexicographically. We then
define the code $x = x_Q = x_T$ by recursion such that for each $n$,
$x(n) =2$ if both $\sigma_n\fr 0$ and $\sigma_n\fr 1$ are in $T$,
$x(n) =1$ if $\sigma_n\fr 0 \notin T$ and $\sigma_n\fr 1 \in T$, and
$x(n) =0$ if $\sigma_n\fr 0 \in T$ and $\sigma_n\fr 1 \notin T$. For a
finite tree $T \subseteq \{0,1\}^{\leq n}$, the finite code $\rho_T$ is
similarly defined, ending with $\rho_T(k)$ where $\sigma_k$ is the
lexicographically last element of $T \cap \{0,1\}^{\leq n}$.

We defined in \cite{BCDW07} a measure $\mu^*$ on the space ${\mathcal
C}$ of closed subsets of $\TN$ as follows.
\begin{equation}
\mu^*({\mathcal X}) = \mu(\{x_Q:Q \in {\mathcal X}\})
\end{equation}
for any ${\mathcal X} \subseteq {\mathcal C}$ and $\mu$ is the
standard measure on $\{0,1,2\}^{\N}$.  Informally this means that
given $\sigma \in T_Q$, there is probability $\frac13$ that both
$\sigma^{\frown}0 \in T_Q$ and $\sigma^{\frown}1 \in T_Q$ and, for
$i=0,1$, there is probability $\frac13$ that only $\sigma^{\frown}i
\in T_Q$. In particular, this means that $Q \cap I(\sigma) \neq
\emptyset$ implies that for $i=0,1$, $Q \cap I(\sigma^{\frown}i) \neq
\emptyset$ with probability $\frac23$.

Then we say that a closed set 
$Q\subseteq 2^{\N}$ is (Martin-L\"{o}f) random if $x_Q$ is
(Martin-L\"{o}f) random.  Note that the equal probability of $\frac13$
for the three cases of branching allows the application of Schnorr's
theorem that \ml randomness is equivalent to prefix-free Kolmogorov
randomness.

The standard (\emph{hit-or-miss}) topology \cite[p.\ 45]{D77} 
on the space $\C$ of closed sets is given by a sub-basis of sets of two types, 
where $U$ is any open set in $2^{\N}$. 
\[
V(U) = \{K: K \cap U \neq \emptyset\}; \qquad \qquad W(U) = \{K: K \subseteq U\}
\]

Note that $W(\emptyset) = \{\emptyset\}$ and that $V(\TN) = \C
\setminus \{\emptyset\}$, so that $\emptyset$ is an isolated element
of $\C$ under this topology.  Thus we may omit $\emptyset$ from $\C$
without complications.

A basis for the hit-or-miss topology may be formed by taking finite
intersections of the basic open sets. We want to work with the
following simpler basis.  For each $n$ and each finite tree $A
\subseteq\{0,1\}^{\leq n}$, let

\[
U_A = \{K\in \C: (\forall \sigma \in \{0,1\}^{\leq n})\  ( \sigma \in A \Iff 
K \cap I(\sigma) \neq \emptyset) \}.
\]
That is, 
\[
U_A = \{K \in \C: T_K \cap \{0,1\}^{\leq n} = A\}.
\] 
Note that the sets $U_A$ are in fact clopen. That is, for any tree $A
\subseteq \{0,1\}^{\leq n}$, define the tree $A' = \{\sigma \in
\{0,1\}^{\leq n}: (\exists \tau \in \{0,1\}^n \setminus A) \sigma
\sqsubseteq \tau\}$. Then $U_{A'}$ is the complement of $U_A$.

For any finite $n$ and any tree $T \subseteq \{0,1\}^{\leq n}$, define the clopen set
$[T] = \bigcup_{\sigma \in T} I(\sigma)$. Then $K \cap [T] \neq \emptyset$
if and only if there exists some $A \subseteq \{0,1\}^{\leq n}$ such that $K
\in U_A$ and $A \cap T \neq \emptyset$. That is, 
\[
V([T]) = \bigcup\{U_A: A \cap T \neq \emptyset\}.
\] 
Similarly, $K \subseteq [T]$
if and only if there exists some $A\subseteq \{0,1\}^n$ such that $K
\in U_A$ and $A \subseteq T$. That is,
\[
W([T]) = \bigcup \{U_A: A \subseteq T\}.
\]
The following lemma can now be easily verified. 

\begin{lem} The family of sets $\{U_A: A \subseteq \{0,1\}^{\leq
    n}\, A\ \text{is a tree}\}$
is a basis of clopen sets for the hit-or-miss topology on $\C$.
\end{lem}

Recall the mapping from $\C$ to $\{0,1,2\}^{\N}$ taking $Q$ to $x_Q$.
It can be shown that this is in fact a homeomorphism.  (See Axon
\cite{Ax10} for details.) Let $\B^*$ be the family of clopen subsets
of $\C$; each set is a finite union of basic sets of the form $U_A$
and thus $\B^*$ is a computable atomless Boolean algebra. Note that
elements $U$ of $\B^*$ are collections of closed sets and are closed
and open in the hit-or-miss topology on the space $\C$ of closed
subsets of $\{0,1\}^{\N}$. Recall that $\B$ denotes the family of
clopen subsets of $\{0,1\}^{\N}$.

\begin{prop} The space $\C$ of nonempty closed subsets of $\TN$ is computably homeomorphic
to the space $\{0,1,2\}^{\N}$. Furthermore, the corresponding map from $\B$ to $\B^*$
is a computable isomorphism of these computable Boolean algebras.   
\end{prop}

Next we consider probability measures $\mu$ on the space $\{0,1,2\}^{\N}$ and the 
corresponding measures $\mu^*$ on $\C$ induced by $\mu$.

A probability measure on $\{0,1,2\}^{\N}$ may be defined as in \cite{RSta} 
from a function $d: \{0,1,2\}^* \to [0,1]$ such that $d(\lambda) = 1$ and,
for any $\sigma \in\{0,1,2\}^*$,
\[
d(\sigma) = \sum_{i=0}^2 d(\sigma \fr i).
\]
The corresponding measure $\mu_d$ on $\{0,1,2\}^{\N}$ is then defined
by letting $\mu_d(I(\sigma)) = d(\sigma)$. Since the intervals
$I(\sigma)$ form a basis for the standard product topology on
$\{0,1,2\}^{\N}$, this will extend to a measure on all Borel sets.  If
$d$ is computable, then $\mu_d$ is said to be computable. The measure
$\mu_d$ is said to be \emph{nonatomic} or \emph{continuous} if
$\mu_d(\{x\}) = 0$ for all $x \in \{0,1,2\}^{\N}$. We will say that
$\mu_d$ is \emph{bounded} if there exist bounds $b,c \in (0,1)$ such
that, for any $\sigma \in \{0,1,2\}^*$ and $i \in \{0,1,2\}$,
\[
b \cdot d(\sigma) < d(\sigma \fr i) < c \cdot d(\sigma).
\]
It is easy to see that any bounded measure must be continuous. We will
say that the measure $\mu_d$ is \emph{uniform} if there exist constants
$b_0,b_1,b_2$ with $b_0+b_1+b_2 = 1$ such that for all $\sigma$ and
for $i \leq 2$, $d(\sigma \fr i) = b_i \cdot d(\sigma)$.

Now let $\mu_d^*$ be defined by \[
\mu_d^*({\mathcal X}) = \mu_d(\{x_Q: Q \in {\mathcal X}\}).
\]
Let us say that a measure $\mu^*$ on $\C$ is computable if the
restriction of $\mu^*$ to the family $\B^*$ of clopen sets is computable. That is, if there is a computable function
$F$ mapping $\B^*$ to $[0,1]$ such that $F(B) = \mu^*(B)$ for all $B \in B^*$. 

\begin{prop} 
For any computable $d$, the measure $\mu^*_d$ is a computable measure on $\C$. 
\end{prop}

\proof  
For any tree $A \subseteq \{0,1\}^{\leq n}$, it is easy to see that

\[
K \in U_A \iff \rho_A \sqsubset x_K, 
\]
so that $\mu_d^*(U_A) = \mu_d(I(\rho_A))$. 
\qed

\section{Computable Capacity and Choquet's Theorem} \label{sec2} 

In this section, we define the notion of capacity and of computable
capacity. We present an effective version of
Choquet's theorem connecting measure and capacity.
For details on capacity and random set
variables, see Nguyen \cite{Ng06} and also Matheron \cite{Ma75}. 

\begin{defi}
A \emph{capacity} on $\C$ is a function $\T: \C \to [0,1]$ with
$\T(\emptyset) =0$ such that
\begin{enumerate}[(1)]
\item $\T$ is monotone increasing, that is, 
\[
Q_1 \subseteq Q_2
  \longrightarrow \T (Q_1) \leq \T(Q_2).
\]
\item $\T$ has the \emph{alternating of infinite order} property, that is,
for $n \geq 2$ and any $Q_1, \dots, Q_n \in \C$
\[
\T(\bigcap_{i=1}^n Q_i) \leq \sum \{(-1)^{|I|+1} \T(\bigcup_{i \in
  I}Q_i): \emptyset \neq I \subseteq \{1,2,\dots,n\} \}.
\]
\item If $Q = \bigcap_n Q_n$ and $Q_{n+1} \subseteq Q_n$ for all
$n$, then $\T(Q) = \lim_{n \to \infty} \T(Q_n)$.
\end{enumerate}
\end{defi}

We will also assume, unless otherwise specified, that the capacity
$\T(\TN) = 1$.

We will say that a capacity $\T$ is computable if it is computable on
the family of clopen sets, that is, if there is a computable function $F$ from 
the Boolean algebra $\B$ of clopen sets into $[0,1]$ such that 
$F(B) = \T(B)$ for any $B \in \B$. 

Define $\T_{d}(Q)=\mu_d^*(V(Q))$.  That is, $\T_{d}(Q)$ is the
probability that a randomly chosen closed set meets $Q$. Here is 
the first result connecting effective measure and effective capacity. 
This follows easily from the classical proof of Choquet. 

\begin{thm} \label{th1}
If $\mu^{*}_{d}$ is a (computable) probability measure on $\C$, then
$\T_{d}$ is a (computable) capacity.
\end{thm}

\proof
Certainly $\T_d(\emptyset) = 0$. The
alternating property follows by basic probability. For (iii), suppose
that $Q = \bigcap_n Q_n$ is a decreasing intersection. Then by
compactness, $Q \cap K \neq \emptyset$ if and only if $Q_n \cap K \neq
\emptyset$ for all $n$. Furthermore, $V(Q_{n+1}) \subseteq V(Q_n)$ for
all $n$. Thus

\[
\T_d(Q) = \mu^*_d(V(Q)) = \mu^*_d(\bigcap_n V(Q_n)) = \lim_n \mu^*_d(V(Q_n)) =
\lim_n\T_d(Q_n).
\]
If $d$ is computable, then $\T_d$ may be computed as
follows.  For any clopen set $I(\sigma_1) \cup \dots \cup
I(\sigma_k)$ where each $\sigma_i \in \{0,1\}^n$, we compute the
probability distribution for all trees of height $n$ and add the
probabilities of those trees which contain one of the
$\sigma_i$. 
\qed

Choquet's Capacity Theorem states that any capacity $\T$ is determined by a measure,
that is $\T = \T_d$ for some $d$. See \cite{Ng06} for details. We now give an 
effective version of Choquet's theorem. It is not so easy, but this does follow from the classical proof of Choquet
\cite{Ch55}. See also \cite{Ma75} and Axon \cite{Ax10}.

\begin{thm} [Effective Choquet Capacity Theorem] \label{th2} 
If $\T$ is a computable capacity, then there is a computable measure
$\mu_d^*$ on the space of closed sets such that $\T =
\T_d$. \end{thm}

\proof
Given the values $\T(U)$ for all clopen sets $I(\sigma_1)\cup \dots
\cup I(\sigma_k)$ where each $\sigma_i \in \{0,1\}^n$, there is in
fact a unique probability measure $\mu_d$ on these clopen sets such
that $\T = \T_d$ and this can be computed as follows.

Suppose first that $\T(I(i)) = a_i$ for $i < 2$ and note that each
$a_i \leq 1$ and $a_0 + a_1 \geq 1$ by the alternating property.  If
$\T = \T_d$, then we must have $d((0)) + d((2)) = a_0$ and $d((1)) +
d((2)) = a_1$ and also $d((0)) + d((1)) + d((2)) = 1$, so that $d((2))
= a_0 + a_1 - 1$, $d((0)) = 1 - a_1$ and $d((1)) = 1 - a_0$. This will
imply that $\T(I(\tau)) = \T_d(I(\tau))$ when $|\tau| = 1$. Now suppose that
we have defined $d(\tau)$ and that $\tau$ is the code for a finite
tree with elements $\sigma_0,\dots,\sigma_n =\sigma$ and thus $d(\tau
\fr i)$ is giving the probability that $\sigma$ will have one or both
immediate successors. We proceed as above. Let $\T(I(\sigma \fr i)) =
a_i \cdot \T(I(\sigma))$ for $i<2$. Then as above $d(\tau \fr 2) =
d(\tau) \cdot (a_0 + a_1 - 1)$ and $d(\tau \fr i) = d(\tau) \cdot (1 -
a_i)$ for each $i$. 
\qed

\section{Zero Capacity}

In this section, we compute the capacity of a random closed set under
certain computable probability measures. In particular, suppose that
$\mu_d$ is a symmetric measure, that is, let $d(\sigma \fr 0) =
d(\sigma \fr 1)$ for all $\sigma$. We show the following. If $d(\sigma
\fr 2) \leq \frac{\sqrt 2}2 d(\sigma)$ for all $\sigma$, then $\T_d(R)
= 0$ for any $\mu_d^*$-random closed set $R$. Thus for the uniform
measure with $d(\sigma \fr 0) = d(\sigma \fr 1) = \frac13 \cdot
d(\sigma)$ for all $\sigma$, effectively random closed sets have
capacity zero. Thus for almost all closed sets $R$, $\T_d(R) = 0$. 
 If $d(\sigma \fr 2) \geq b \cdot d(\sigma)$ for all
$\sigma$, where $b > \frac{\sqrt 2}2$ is a constant, then $\T_d(R) >
0$ for any $\mu^*_d$-random closed set $R$. Thus for almost all closed
sets $R$, $\T_d(R) > 0$. This result, and others in this section are
new results about classical measure and capacity as well as results about
algorithmic randomness. 
 
For non-symmetric measures, where $d(\sigma \fr i) = b_i \cdot d(\sigma)$ for $i < 2$,
the question of whether a random closed set has zero capacity depends
on the sum $b_0+b_1$ and also on their difference. If $b_0 + b_1 \geq
2 - \sqrt 2$ and $|b_0 - b_1|$  is
sufficiently small, then every $\mu_d^*$-random closed set will have
capacity zero (so that for almost all closed sets $R$, $\T_d(R) =
0$) and otherwise there is a $\mu_d^*$-random closed set
with positive capacity. 

We say that $K \in \C$ is \emph{$\mu_d^*$-random} if $x_K$ is \ml random
with respect to the measure $\mu_d$. (See \cite{RSta} for details.)
Our results show that the $\T_d$ capacity of a $\mu^*_d$-random closed set 
depends on the particular measure.

In the following proofs, the key idea is that an arbitrary closed set
$Q$ can be given as the intersection of a sequence $\left< Q_n
\right>_{n \in \omega}$ of clopen sets, so that the
capacity $\T(Q) = \lim_n \T(Q_n)$. Thus we want to compute the capacity
$q_n$ of $Q_n$ when $Q$ is a random closed set, or at least to compute
bounds on this capacity. Now the capacity of $Q$ is the probability
that $Q \cap K \neq \emptyset$ for a random closed set $K$, that is to
say  $\T(Q) = \mu^*_d(\{K: Q \cap K \neq \emptyset\})$. Thus we first
compute the probability that $Q_n \cap K_n \neq \emptyset$ for randomly
chosen closed sets $Q$ and $K$ and use this to determine $\T(Q)$ for a
random closed set. In the first two theorems, these computations can
be converted into \ml tests, so that the capacity of an effectively
$\mu_d^*$-random closed set can be determined. 

\begin{thm}\label{th4a} 
Suppose that the measure $\mu_d$ is defined by $d$ such that, for all
sufficiently long $\sigma \in \{0,1\}^*$, $d(\sigma \fr 2) \leq
\frac{\sqrt 2}2 d(\sigma)$ and $d(\sigma \fr 0) = d(\sigma \fr
1)$. Then, for any $\mu_d^*$-random closed set $R$, $\T_d(R) = 0$. 
Thus for almost all closed sets $R$, $\T_d(R) = 0$. 
\end{thm}

\proof
We first present the proof for a uniform measure $\mu_d$ and then give
the modifications necessary for non-uniform measure. 

Fix $b$ with $1 - 2b \leq \frac{\sqrt 2}2$ and suppose that, for all
$\sigma$, $d(\sigma \fr 2) = (1 - 2b) \cdot d(\sigma)$ and, for
$i=0,1$, $d(\sigma \fr i) = b \cdot d(\sigma)$. Now let $\mu^* =
\mu_d^*$. We will compute the probability, given two closed sets $Q$
and $K$, that $Q \cap K$ is nonempty.  Here we define the usual
product measure on the product space $\C \times \C$ of pairs $(Q,K)$
of nonempty closed sets by letting $\mu^2(U_A \times U_B) = \mu^*(U_A)
\cdot \mu^*(U_B)$ for arbitrary subsets $A,B$ of $\{0,1\}^n$.

Let
\[
Q_n = \bigcup \{I(\sigma): \sigma \in \{0,1\}^n\ \&\ Q \cap I(\sigma)\neq \emptyset\}
\]
and similarly for $K_n$. Then $Q \cap K \neq \emptyset$ if and only if
$Q_n \cap K_n \neq \emptyset$ for all $n$. Let $p_n$ be the
probability that $Q_n \cap K_n \neq \emptyset$ for two arbitrary
closed sets $K$ and $Q$, relative to our measure $\mu^*$. It is
immediate that $p_1 = 1 - 2b^2$, since $Q_1 \cap K_1 = \emptyset$ only
when $Q_1 = I(i)$ and $K_1 = I(1-i)$. Next we will determine the
quadratic function $f$ such that $p_{n+1} = f(p_n)$. There are 9
possible cases for $Q_1$ and $K_1$, which break down into 4 distinct
cases in the computation of $p_{n+1}$. 

\begin{desCription}

\item\noindent{\hskip-12 pt\bf Case (i):}\
  As we have seen, $Q_1 \cap K_1 = \emptyset$ with
  probability $1 - 2b^2$.\medskip

\item\noindent{\hskip-12 pt\bf Case (ii):}\
  There are two chances that $Q_1 = K_1 = I(i)$, each
  with probability $b^2$ so that $Q_{n+1} \cap K_{n+1} \neq \emptyset$
  with probability $p_n$.\medskip

\item\noindent{\hskip-12 pt\bf Case (iii):}\
  There are four chances where $Q_1 = \TN$ and $K_1
=I(i)$ or vice versa, each with probability $b \cdot (1-2b)$, so that
once again $Q_{n+1} \cap K_{n+1} \neq\emptyset$ with relative probability
$p_n$.\medskip

\item\noindent{\hskip-12 pt\bf Case (iv):}\
 There is one chance that $Q_1 = K_1 = \TN$, with
probability $(1 - 2b)^2$, in which case $Q_{n+1} \cap K_{n+1} \neq
\emptyset$ with relative probability $1 - (1 -p_n)^2 = 2p_n -
p_n^2$. This is because $Q_{n+1} \cap K_{n+1} = \emptyset$ if and only
if both $Q_{n+1} \cap I(i) \cap K_{n+1} = \emptyset$ for both $i=0$
and $i=1$.\medskip

\end{desCription}

\noindent Adding these cases together, we see that 
\[
p_{n+1} = [2b^2 + 4b (1-2b)] p_n + (1 - 2b)^2 (2p_n - p_n^2) = (2b^2 -
4b + 2) p_n  - (1 - 4b + 4b^2) p_n^2. 
\]

Next we investigate the limit of the computable sequence $\left< p_n \right>_{n \in
 \omega}$. Let $f(p) =  (2b^2 - 4b + 2) p  - (1 - 4b + 4b^2) p^2$. 
Note that $f(0) = 0$ and $f(1) = 1 - 2b^2 < 1$. 
It is easy to see that the fixed points of $f$ are $p=0$ and $p =
 \frac{2b^2 -4b+1}{(1-2b)^2}$. Note that since $b < \frac 12$, the
 denominator is not zero and hence is always positive.  

Now consider the function $g(b) = 2b^2 - 4b +1 = 2 (b-1)^2 - 1$, which has
positive root $\ob = 1 - \frac{\sqrt 2}2$ and is decreasing for $0 \leq b \leq 1$.

There are three cases to consider
when comparing $b$ with $\ob$.

\begin{desCription}

\item\noindent{\hskip-12 pt\bf Case 1:}\ If $b > \ob$, then $g(b) < 0$
  and hence the other fixed point of $f$ is negative. Furthermore,
  $2b^2 - 4b +2 < 1$ so that $f(p) < p$ for all $p > 0$. It follows
  that the sequence $\{p_n: n \in \N\}$ is decreasing with lower bound
  zero and hence must converge to a fixed point of $f$ (since $p_{n+1}
  = f(p_n)$). Thus $\lim_n p_n = 0$.

\item\noindent{\hskip-12 pt\bf Case 2:}\ If $b = \ob$, then $g(b) = 0$
  and $f(p) = p - (4b-1) p^2$, so that $p=0$ is the unique fixed point
  of $f$. Furthermore, $4b-1 = 3 - 2 \sqrt2 > 0$, so again $f(p) < p$
  for all $p$. It follows again that $\lim_n p_n = 0$.

\end{desCription}

\noindent In these two cases, we can define a \ml test to prove that $T_d(R) =
0$ 
for any $\mu$-random closed set $R$. 

For each $m, n \in \N$, let 
\[
B_m = \{(K,Q): K_m \cap Q_m \neq \emptyset\},
\]
so that $\mu^*(B_m) = p_m$ and let
\[
A_{m,n} = \{Q: \mu^*(\{K: K_m \cap Q_m \neq \emptyset\}) \geq
2^{-n}\}.
\]

\begin{clm}\label{c1} 
  For each $m$ and $n$, $\mu^*(A_{m,n}) \leq 2^n \cdot p_m$. 
\end{clm}

\proof Define the Borel measurable function $F_m: \C \times \C \to
\{0,1\}$ to be the characteristic function of $B_m$. Then
\[
p_m = \mu^2(B_m) = \int_{Q \in \C} \int_{K \in \C} F(Q,K) dK dQ.
\]
Now for fixed $Q$, 
\[
\mu^*(\{K: K_m \cap Q_m \neq \emptyset\}) = \int_{K \in \C} F(Q,K) dK,
\]
so that for $Q \in A_{m,n}$, we have $\int_{K \in \C} F(Q,K) dK \geq 2^{-n}$. 
It follows that 
\begin{align*}
p_m = \int_{Q \in \C} \int_{K \in \C} F(Q,K) dK dQ &\geq \int_{Q \in
  A_{m,n}} \int_{K \in \C} F(Q,K) dK dQ \\
&\geq \int_{Q \in A_{m,n}}
2^{-n} dQ = 2^{-n} \mu^*(A_{m,n}).
\end{align*}
Multiplying both sides by $2^n$ completes the proof of Claim \ref{c1}. 
\qed

\medskip

Since the computable sequence $\langle p_n\rangle_{n \in \omega}$ converges to 0,
there must be a computable subsequence $m_0,m_1,\dots$ such that
$p_{m_n} < 2^{-2n-1}$ for all $n$. We can now define our \ml test. Let

\[
S_r = A_{m_r,r}
\]
and let
\[
V_n = \bigcup_{r>n} S_r.
\]
It follows that
\[
\mu^*(S_r) \leq 2^{r+1}\mu^*(B_{m_r}) < 2^{r+1} 2^{-2r-1} = 2^{-r}
\]
and therefore 
\[
\mu^*(V_n) \leq \sum_{r>n} 2^{-r} = 2^{-n}
\]
Now suppose that $R$ is a random closed set. The sequence $\la V_n
\ra_{n \in \omega}$ is a computable sequence of c.~e. open sets with
measure $\leq 2^{-n}$, so that there is some $n$ such that $R \notin
S_n$. Thus for all $r > n$, $\mu^*(\{K: K_{m_r} \cap R_{m_r} \neq
\emptyset\}) < 2^{-r}$ and it follows that
\[
\mu^*(\{K: K \cap R \neq \emptyset\}) = \lim_n \mu^*(\{K: K_{m_n} \cap
R_{m_n} \neq \emptyset\}) = 0.
\]
Thus $\T_d(R) = 0$, as desired. 

This completes the proof when the function $d$ is independent of $\sigma$.

\bigskip

Next suppose that the value $b$ such that $d(\sigma \fr i) = b \cdot
d(\sigma)$ for $i=0,1$, depends on $\sigma$, say $b_{\sigma} = d(\sigma \fr
i) / d(\sigma)$ and that $b_{\sigma} \geq \ob$ for all $\sigma$. 

Let $f_b(p) = (2b^2
- 4b + 2) p - (1 - 4b + 4b^2) p^2$ as above and let $f_{\ob}(p) = f(p)$. Let
$p_n$ be the probability computed above corresponding to $b_{\sigma} =
\ob$ for all $\sigma$, so that $p_{n+1} = f(p_n)$. Define $p_n^d$ to
be the probability, under $\mu_d^*$, that $K_n \cap Q_n \neq
\emptyset$, for closed sets $K$ and $Q$. We will argue by induction on $n$ 
that $p^d_n \leq p_n$.

\medskip

\begin{clm}\label{c2} 
For any reals $b,c,p \in [0,1]$, if $b < c$, then $f_c(p) \leq f_b(p)$.
\end{clm}

\proof Fixing $p$ and taking the derivative of $f_b(p)$ with respect to $b$,
we obtain 
\[
\frac{\partial f}{\partial b}(b,p) = (4b-4)p - (8b-4)p^2 \leq -4bp \leq 0,
\]
with the inequality due to the fact that $p^2 \leq p$ on $[0,1]$.
\qed

Now suppose that for all $\sigma \in \{0,1\}^*$ and for $i < 2$,
$d(\sigma \fr i) \geq \ob d(\sigma)$ and again let $p_n^d$ be the 
$\mu_d$-probability that $K_n \cap Q_n \neq \emptyset$.
Clearly $p_0^d = 1 = p_0$.

Now assume that $p_n^d \leq p_n$ for any $d$ as above. Let $d$ be given as above 
with $d((0)) = d((1)) = b \geq \ob$ and define $d_i$ for $i=0,1$ as follows.
\[
d_i(\sigma \fr j) = d(i \fr \sigma \fr j).
\]
Let $p^i$ be the probability under $d_i$ that $Q_n \cap K_n \neq \emptyset$.  
Then the probability under $d_{i+1}$ that $Q_{n+1} \cap K_{n+1} \neq \emptyset$ can be 
computed in the four cases as above to equal
\[
b^2(p^0 + p^1) + 2b(1-2b) (p^0 + p^1) + (1 - 2b)^2 (1 - (1-p^0)(1-p^1)).
\]
By induction, both of $p^0$ and $p^1$ are $\leq p_n$ and it follows easily that
\[
b^2(p^0 + p^1) + 2b(1-2b) (p^0 + p^1) + (1 - 2b)^2 (1 -
(1-p^0)(1-p^1)) \leq f_b(p_n) \leq f(p_n) = p_{n+1}.
\]
Finally, suppose that we only have that $b_{\sigma} \geq \ob$ for
$\sigma$ with $|\sigma| \geq n$.  Let $R$ be $\mu_d^*$-random and for
each $\sigma$ of length $n$, let $d_{\sigma}$ be defined so that
$d_{\sigma}(\tau) = d(\sigma \fr \tau)$ and let $R_{\sigma} = \{X:
\sigma \fr X \in R\}$. Then $R_{\sigma}$ is $d_{\sigma}$-random for
each $\sigma$, so that the capacity $\T_{d_{\sigma}}(R_{\sigma}) = 0$.  It
follows that $\T_d(R) = 0$ since $Q \cap R \neq \emptyset$ if and only
if $Q \cap R \cap I(\sigma) \neq \emptyset$ for some $\sigma$ of
length $n$.

The appropriate \ml test can now be given as before to show that any $\mu_d^*$-random closed set
will have capacity zero. 
\qed

Next we consider the case where random closed sets will have positive capacity.

\begin{thm}\label{th4b} 
Suppose that $b < \ob = 1 - \frac{\sqrt 2}2$ is fixed and that the
measure $\mu_d$ is defined by $d$ such that, for all sufficiently long
$\sigma$, $d(\sigma \fr 0) = d(\sigma \fr 1) \leq b \cdot d(\sigma)$.  
Then $\{R \in \C: \T_d(R) > 0\}$ has $\mu_d^*$ measure one and
furthermore every $\mu_d^*$-random closed set has positive capacity.
Thus for almost all closed sets $R$, $\T_d(R) > 0$. 
\end{thm}

\proof
First fix $b < \ob$ and fix $d$ so that $d(\sigma
\fr i) = d(\sigma) \cdot b$ for all $\sigma$ and for $i < 2$, and let $\mu^* = \mu_d^*$. 
Since $0 < 2b^2 - 4b +1 < 1$, the function $f = f_b$ defined above has a positive fixed point
$m_b = \frac{2b^2 -4b+1}{(1-2b)^2}$.  It is clear that $f(p) > p$ for
$0 < p < m_b$ and $f(p) < p$ for $m_b < p$. Furthermore, the function
$f$ has its maximum at $p = [\frac{1-b}{1-2b}]^2 > 1$, so that $f$ is
monotone increasing on $[0,1]$ and hence $f(p) > f(m_b) = m_b$
whenever $p > m_b$.  As in the proof of Theorem \ref{th4a} let $p_n$ be the probability
that $Q_n \cap K_n \neq \emptyset$ for arbitrary closed sets $Q$ and $K$. 
Observe that $p_0 = 1 > m_b$ and hence the
sequence $\{p_n: n \in \N\}$ is decreasing with lower bound $m_b$. It
follows that $\lim_n p_n = m_b > 0$.

Now $B = \{(Q,K): Q \cap K \neq \emptyset\} = \bigcap_n B_n$ is the
intersection of a decreasing sequence of sets and hence $\mu^2(B) =
\lim_n p_n = m_b >0$.

\begin{clm}\label{nc2} $\mu^*(\{Q: \mu^*(\{K: K \cap Q \neq
\emptyset\}) > 0\}) \geq m_b$.
\end{clm}

\proof Let $B = \{(K,Q): K \cap Q \neq \emptyset$,
let $A = \{Q: \mu^*(\{K: K \cap Q \neq \emptyset\}) > 0\}$ and suppose
that $\mu^*(A) < m_b$.  As in the proof of Claim \ref{c1}, we have
\[
m_b = \mu^2(B) = \int_{Q \in \C} \int_{K \in \C} F(Q,K) dK dQ.
\]
For $Q \notin A$, we have $\int_{K \in Q} F(Q,K) dK = \mu^*(\{K: K
\cap Q \neq \emptyset\}) = 0$, so that
\[
m_b = \int_{Q \in A} \int_{K \in Q} F(Q,K) dK dQ \leq \int_{Q \in A} dQ = \mu^*(A),
\]
which completes the proof of Claim \ref{nc2}. \qed

\begin{clm}\label{nc3} $\{Q: \T_d(Q) \geq m_b\}$ has positive measure. 
\end{clm}

\proof Recall that $T_d(Q) = \mu^*(\{K: Q \cap K \neq \emptyset\})$.
Let $B = \{(K,Q): K \cap Q \neq \emptyset$, let $A = \{Q: T_d(Q) \geq m_b\}$ 
and suppose that $\mu^*(A) = 0$. As
in the proof of Claim \ref{c1}, we have
\[
m_b = \mu^2(B) = \int_{Q \in \C} T_d(Q) dQ.
\]
Since $\mu^*(A) = 0$, it follows that for any $B \subseteq \C$, we have
\[
\int_{Q \in B} T_d(Q) dQ \leq m_b \mu^*(B).
\]
Furthermore, $T_d(Q) < m_b$ for almost all $Q$, so there exists some $P$ with
$T_d(P) < m_b - \epsilon$ for some positive $\epsilon$. This means that for some
$n$, $\mu^*(\{K: P_n \cap K_n \neq \emptyset\}) < m_b - \epsilon$. Then for \emph{any}
closed set $Q$ with $Q_n = P_n$, we have $T_d(Q) < m_b - \epsilon$. But 
$E = \{Q: Q_n = P_n\}$ has positive measure, say $\delta > 0$. Then we have

\begin{align*}
m_b = \int_{Q \in \C} T_d(Q) dQ &= \int_{Q \in E} T_d(Q) dQ\ +\ \int_{Q \notin E} T_d(Q) dQ \\
&\leq \ \delta (m_b - \epsilon) + (1- \delta) m_b = m_b - \epsilon \delta < m_b.
\end{align*}
This contradiction demonstrates Claim \ref{nc3}. \qed

It is now easy to see that $\T_d(R) > 0$ with probability one.
That is, let $p$ be the probability that $\T_d(R) = 0$. Then by considering the first
level of $R$, we can see that $p = 2bp + (1 - 2b)p^2$ and hence either $p=0$ or $p=1$.
Since we know that $p < 1$, it follows that $p=0$.

Since the set of $\mu^*$-random closed sets has measure one, there must
be a random closed set $R$ such that $\T_d(R) \geq m_b$ and
furthermore, almost every $\mu^*$-random closed set has positive
capacity.

Furthermore, we can construct a \ml test as follows. First observe that for any
computable $q$, $\{Q: \T_d(Q) < q\}$ is a c.~e.\ open set. This is because $\T_d(Q) < q \iff
(\exists n) \T_d(Q_n) < q$ and $\T_d(Q_n)$ can be uniformly computed from $Q$. 
 
Now let $h(p)$ be the probability that $\T_d(Q) < p$. Note that if
$\T_d(Q_i) \geq p$ for $i=0$ or for $i=1$, then $\T_d(Q) \geq bp$. It
follows that $h(bp) \leq h(p)^2$.  Since $\T_d(Q) = 0$ with
probability zero, it follows that $\lim_{p \to 0} h(p) = 0$.  Take a
rational $q$ small enough so that $h(q) < \frac12$.  Then $h(b^n q)
\leq (\frac12)^{2^n} \leq 2^{-n}$. Let $S_n = \{Q: \T_d(Q) \leq b^n
q\}$.  Then $\mu_d^*(S_n) \leq 2^{-n}$ and the sequence $(S_n)$ is
effectively c.~e. open, so that no random closed set can be belong to
all $S_n$.  But if $\T_d(Q) = 0$, then of course $Q \in S_n$ for all
$n$. Thus every $\mu_d^*$ random closed set must have positive
capacity.

This completes the proof when $d$ is independent of $\sigma$. 

Next suppose that $b < \ob$ and that, for all $\sigma$, $d(\sigma \fr
0) = d(\sigma \fr 1) \leq b \cdot d(\sigma)$. Let $p^d_n$ now be the $\mu_d^*$ probability that 
$Q_n \cap K_n \neq \emptyset$. It follows from the monotonicity of $f$ (Claim \ref{c2})
that $p^d_n \geq p_n$ for all $d$ as above and thus $\lim_n p^d_n \geq m_b$. 
The same argument as above now shows that $\{Q: \T_d(Q) \geq m_b\}$ has positive measure
and thus $\T_d(Q)$ has positive capacity with probability one. The argument that
every $\mu_d^*$-random closed set has positive capacity follows as above. 
\qed

Note that random closed sets can have arbitrarily small positive
capacity. This follows from the fact that $\T_d(0 \fr Q) = (1-b) \T_d(Q)$. 

Thus for certain measures, there exists a random closed set with
measure zero but with positive capacity. For the standard measure, a random closed set
has capacity zero. 

\begin{cor} 
\label{th3} 
Let $d$ be the uniform measure with $b_0 = b_1 = b_2 = \frac13$. Then
for any $\mu_d^*$-random closed set $R$, $\T_d(R) = 0$. \qed
\end{cor}

Finally, we consider non-symmetric measures, where $d(\sigma \fr 0)$
does not necessarily equal $d(\sigma \fr 1)$. We will give the result
where $\mu_d$ is a uniform measure. The proofs follow the same outline
as those of Theorems \ref{th4a} and \ref{th4b}. 

\begin{thm} \label{th4c} Fix $b$ and let $\mu_{d}$ be a measure defined by $d$ where 
$d(\sigma \fr i) = b_i \cdot d(\sigma)$ with $b_{0}+b_{1}=2b>0$ and
  $b_{2}=1-2b>0$ and let $\hat{b}=1-\frac{\sqrt{2}}{2}$. Then
\begin{enumerate}[\em(1)]
\item If $b\geq \hat{b}$ and
  $|b_{0}-b_{1}| \leq \sqrt{8b-4b^{2}-2}$, then for any
  $\mu^{*}_{d}$-random closed set $R$, $\T_{d}(R)=0$.
Thus for almost all closed sets $R$, $\T_d(R) = 0$. 

\item If $b>\hat{b}$ or $|b_{0}-b_{1}| > \sqrt{8b-4b^{2}-2}$, then there 
is a $\mu_{d}^{*}$-random closed set $R$ with $\T_{d}(R)>0$.
	\end{enumerate}
\end{thm}

\proof For convenience let $\mu=\mu^{*}_{d}$ and let $\mu^{2}=\mu\times\mu$ be the usual 
product measure on the product space $\C \times \C$. We will compute
the probability $p = \mu^{2}(\{(Q,K):Q \cap K \neq \emptyset\})$.

 As in the proof of Theorem \ref{th4a} let $p_n$ be the probability
 that $Q_n \cap K_n \neq \emptyset$ for arbitrary closed sets $Q$ and
 $K$, so that $p = \lim_n p_n$ Clearly,
 $p_{1}=1-2b_{0}b_{1}$ since $Q_{1}\cap K_{1} =\emptyset$ only when
 $Q_{1}=I(i)$ and $K_{1}=I(1-i)$. We will compute as before a
 quadratic function $f$ so that $p_{n+1}=f(p_{n})$.  Considering the
 various cases as in the proof of Theorem \ref{th4a}, we see that
		
\begin{align*}
	p_{n+1}&=(b_{0}^{2}+b_{1}^{2}+4b(1-2b))p_{n}+(1-2b)^{2}(2p_{n}-p_{n}^{2})\\
	&=(2b_{0}-4bb_{0}+4b^{2}+4b+2)p_{n}-(1-2b)^{2}p_{n}^{2}
\end{align*}
		
Next, we investigate $\lim_{n} p_{n}$. Let 
\[
f(p)=(2b_{0}-4bb_{0}+4b^{2}+4b+2)p-(1-2b)^{2}p^{2}
\]
This function has fixed points $p=0$ and
$p=\frac{2b_{0}-4bb_{0}+4b^{2}+4b+1}{(1-2b)^{2}}$. Note that we must
have $b<\frac{1}{2}$ so $(1-2b)^{2}>0$.

Now consider the functions $g(a)=2a-4ba+4b^{2}+4b+1$, which has roots
$a_{\pm}=b\pm\sqrt{-b^{2}+2b-\frac{1}{2}}$ and
$h(b)=-b^{2}+2b-\frac{1}{2}=-2\big(2(b-1)^{2}-1\big)$, which has root
$\hat{b}$. There are 3 cases to consider when comparing $b$ and
$\hat{b}$.
\begin{enumerate}[(1)]
\item If $b>\hat{b}$ and $a_{-}\leq b_{0}\leq a_{+}$, then
  $g(b_{0})<0$ and hence the nonzero fixed point of $f$ is
  negative. Since $(p_{n})$ is decreasing with lower bound $0$ the
  sequence converges to a non-negative fixed point of $f$. Hence
  $p = \lim_{n}p_{n}=0$.

\item If $b=\hat{b}$ and $b_{0}=b$ or if $b_{0}=a_{\pm}$ then
  $g(b_{0})=0$ and so $p=0$ is the only fixed point of $f$ hence
  $p = \lim_{n}p_{n}=0$.

\item If $b < \hat{b}$ or $b_{0}\not \in [a_{-},a_{+}]$, then
  $g(b_{0})>0$ and so $f$ has positive fixed point
  $m_{b,b_{0}}=\frac{2b_{0}-4bb_{0}+4b^{2}+4b+1}{(1-2b)^{2}}$. Furthermore,
  $f$ has its maximum at
  $p=\frac{b_{0}-2bb_{0}+2b^{2}+2b+1}{(1-2b)^{2}}>1$ (since
  $2b>2bb_{0}$). Thus $f$ is increasing for $p<1$, so if
  $p>m_{b,b_{0}}$, then $f(p)>f(m_{b,b_{0}})=m_{b,b_{0}}$. Hence,
  since $p_{0}=1$, $(p_{n})$ is bounded below by $m_{b,b}$ and so,
  $p = \lim_{n}p_{n}=m_{b,b_{0}}>0$. \qed
\end{enumerate}	

\noindent Due to he inequalities needed for $|b_0-b_1|$ in the theorem, it seems
that the proof given above does not easily extend to provide a result
for non-uniform measures or to prove that, in the second case above,
\emph{every} random closed set has positive capacity. 

\section{Effectively Closed Sets}

In this section, we consider the capacity of effectively closed sets. 
A random closed set can never be effectively closed. But we can still 
construct an effectively closed set with measure zero and with
positive capacity. 

We begin by characterizing the possible capacity of effectively closed
sets.  For the following results we will take $\T=\T_d$ where $\mu_d$ is the
computable measure defined by $d(\sigma \fr i) = b_i$ with $0< b_1 \leq
b_0$ and $1 > b_0 + b_1 > 0$. For any effectively closed set $Q = [T]$, $Q$ is the effective
intersection of the decreasing sequence $[T_n]$ of clopen sets,
where $T_n = T \cap \{0,1\}^{\leq n}$. Thus for a computable measure $\T_d$, 
the capacity $\T_d(Q)$ is the limit of a computable, decreasing sequence and is therefore
an upper semi-computable real. We will show that for \emph{every} upper semi-computable
real $q \in [0,1]$, there exists an effectively closed set $Q$ with $\T_d(Q) = q$. 

\begin{lem} Let $Q= 0\fr Q_0 \cup 1\fr Q_1$ and let  $q_i = \T(Q_i)$ for $i \leq 1$.
Then, $\T(Q)=(1-b_1)q_0 + (1-b_0)q_1 - (1-(b_0 + b_1)) \cdot q_0 q_1$.
\end {lem}

\proof
For a closed set $K$, $K \cap Q\neq\emptyset$ if and only if one of the following holds:
\begin{enumerate}[(1)]
\item $K=0\fr K_{0}$ and $Q_{0} \cap K_{0}\neq \emptyset$ 
(which has probability $b_0 \cdot \T(Q_{0})$), or
\item $K=1\fr K_{1}$ and $Q_{1} \cap K_{1}\neq \emptyset$ 
(which has probability $b_1 \cdot \T(Q_{1})$), 
or
\item $K= 0 \fr K_{0}\cup 1 \fr K_{1}$ and either $Q_{0} \cap K_{0} \neq \emptyset$ 
or $Q_{1} \cap K_{1}\neq \emptyset$ (which has probability 
$(1-(b_0 +b_1)) (1-(1-\T(Q_0)(1-\T(Q_1))$).
\end{enumerate}
Thus, 
	\begin{align*}
	\T(Q)&= b_0 q_0 + b_1q_1 +(1-(b_0+b_1))(1-(1-q_0)(1-q_1))\\
	&=(1-b_1)q_0 + (1-b_0)q_1 -(1-(b_0 + b_1))q_0 q_1
		\end{align*} \qed

\begin{lem}\label{stop} Let $Q= \bigcup_{k=0}^{k=n}I(\sigma_k)$. 
Then for each $j \leq k$, $\T(Q)-\T(Q \setminus I(\sigma_j)) \leq (1-b_1)^{|\sigma_j|}$.
\end{lem}

\proof The proof is by induction on $|\sigma_j|$. If $|\sigma_j| = 0$,
this is trivial.  

Let $Q = 0 \fr Q_0 \cup 1 \fr Q_1$ and let $q_i
=\T(Q_i)$ for $i=0,1$.  If $\sigma_i = (i)$, then $\T(Q) = (1-b_{1-i})+
b_{1-i} \cdot q_i$ and $\T(Q \setminus I(i))=(1-b_i) \cdot q_{1-i}$.  Thus, $\T(Q)
- T(Q \setminus I(i))= (1-b_{1-i}) - (1-(b_0+b_1)) \cdot q_i \leq 1-b_1$.
	
Now let $|\sigma_j|=n > 0$ and let $\sigma_j= i \fr \tau$ for some $i
\leq 1$ and some $\tau$.  Let $r = \T(Q_i \setminus I(\tau))$. Then,
$\T(Q)-\T(Q \setminus I(\sigma_j))= (1-b_{1-i}) (q_i-r) -
(1-(b_0 + b_1)) q_{1-i} (q-r) \leq (1-b_1)(q-r) \leq (1-b_1)(1-b_1)^{n-1}$, where the
last inequality holds by the induction hypothesis.
\qed

\begin{thm} \label{th5} Let the real number $q \in [0,1]$ be upper semi-computable, 
i.e.\ there is a computable, decreasing sequence $\{q_n: n \in \N\}$
such that $\lim q_n =q$. Then there exists an effectively closed set $P$ such that
$\T(P)=q$. Moreover, $P$ can be written as $\bigcap_{n} P_n$ where
$\{P_n: n \in \N$ is a computable sequence of clopen sets with
$q_{n+1} \leq \T(P_n) \leq q_{n}$.
\end{thm}

\proof  We may assume without loss of generality that $q_0 = 1$. 
We will construct $P_n$ by recursion beginning with $P_0 =
2^{\N}$.  Now suppose we have constructed the clopen set $Q_{n-1} =
\bigcup_{k=0}^m I(\sigma_k)$ such that $q_n \leq \T_d(Q_{n-1}) \leq
q_{n-1}$.

Let $\delta = q_n - q_{n-1}$ and compute $s$ large enough so that 
$(1-b_1)^s < \delta$ and $|\sigma_k| \leq s$ for all $k \leq m$. 
Then we can rewrite each interval $I(\sigma_k)$ as a union of intervals $I(\tau)$ 
with $|\sigma| = s$ and thus obtain $Q_{n-1} = \bigcup_{k=0}^r I(\tau_k)$ with 
$|\tau_k| = s$ for all $k \leq r$. Now let $Q_{n-1,k} = \bigcup_{j=0}^{k-1} I(\tau_j)$ for each 
$k \leq r+1$, so that $Q_{n-1,k} = Q_{n-1,k+1} \setminus I(\tau_k)$ for each $k \leq r$. 
Observe that $\T_d(Q_{n-1,r+1}) = \T_d(Q_{n-1}) \geq q_n$ and that 
$\T_d(Q_{n-1,0}) =\T_d(\emptyset) = 0 \leq q_n$. 

It follows from Lemma \ref{stop} that, for any $k$, $\T_d(Q_{n-1,k+1}) -
\T_d(Q_{n-1,k}) \leq \delta$. Now let $k$ be the least such that $\T_d(Q_{n-1,k}) \leq
q_n$. Then $\T_d(Q_{n-1,k+1}) > q_n$ and also $\T_d(Q_{n-1,k+1}) \leq \T_d(Q_{n-1,k}) + \delta
\leq q_n + \delta \leq q_{n-1}$. So we let $Q_n = Q_{n-1,k+1}$. 
	
In this way, we have constructed a computable, decreasing sequence
$Q_n$ of clopen sets with $q_n \leq \T_d(Q_n) \leq q_{n-1}$, so that,
for $Q = \bigcap_nQ_n$, we have $\T_d(Q) = \lim_n \T_d(Q_n)= q$. \qed

\begin{thm}  \label{th6}
For the uniform measure $\mu_d$ defined by $d(\sigma \fr i) = b \cdot
d(\sigma)$ for all $\sigma$, there is an effectively closed set $Q$ with
Lebesgue measure zero and positive capacity $\T_d(Q).$
\end{thm}

\proof
First let us compute the capacity of $X_n = \{x: x(n) =0\}$. For
$n=0$, we have $\T_d(X_0) = 1 - b$. That is, $Q$ meets $X_0$ if and
only if $Q_0 = I(0)$ (which occurs with probability $b$), or $Q_0 =
\TN$ (which occurs with probability $1 - 2b$). Now the probability
$\T_d(X_{n+1})$ that an arbitrary closed set $K$ meets $X_{n+1}$ may
be calculated in two distinct cases.  As in the proof of Theorem
\ref{th3}, let
\[
K_n = \bigcup \{I(\sigma): \sigma \in \{0,1\}^n\ \&\ K \cap I(\sigma)\neq \emptyset\}
\]

\begin{desCription}
\item\noindent{\hskip-12 pt\bf Case I:}\
  If $K_0 = \TN$, then $\T_d(X_{n+1}) = 1 - (1- \T_d(X_n))^2$.\medskip

\item\noindent{\hskip-12 pt\bf Case II:}\
  If $K_0 = I((i))$ for some $i<2$, then $\T_d(X_{n+1}) = \T_d(X_n)$.\medskip 

\end{desCription}

\noindent It follows that 
\begin{align*}
\T_d(X_{n+1}) &= 2b \cdot \T_d(X_n) + (1-2b) (2 \T_d(X_n)
- (\T_d(X_n))^2) \\
&= (2-2b) \T_d(X_n) - (1-2b) (\T_d(X_n))^2
\end{align*}

Now consider the function $f(p) = (2-2b) p - (1-2b) p^2$, where $0 < b
 < \frac 12$. This function has the properties that $f(0) = 0$, $f(1) =
 1$ and $f(p) > p$ for $0 < p < 1$. Since $\T_d(X_{n+1}) 
= f(\T_d(X_n))$, it follows that $\lim_n \T_d(X_n) = 1$ and is the limit
of a computable sequence.

For any $\sigma = (n_0, n_1, \dots, n_k) \in \N^{\N}$, with $n_0 < n_1
< \cdots <n_k$, similarly define $X_{\sigma} = \{x: (\forall i \leq k)
x(n_i) = 0\}$.  A similar argument to that above shows that $\lim_n
\T_d(X_{\sigma \fr n}) / \T_d(X_{\sigma}) = 1$.

Now consider the decreasing sequence $c_k = \frac{2^{k+1} +
1}{2^{k+2}}$ with limit $\frac12$.  Choose $n = n_0$ such that
$\T_d(X_n) \geq \frac 34 = c_0$ and for each $k$, choose $n = n_{k+1}$
such that $\T_d(X_{(n_0,\dots,n_k,n)}) \geq c_{k+1}$. This can be done
since $c_{k+1} < c_k$. Finally, let $Q = \bigcap_k
X_{(n_0,\dots,n_k)}$.  Then $\T_d(Q) = \lim_k \T_d(X_{(n_0,\dots,n_k)})
\geq \lim_k c_k = \frac12$.  \qed

It is clear that we can make the capacity in Theorem \ref{th6} arbitrarily large
below 1.  

\section{Conclusions}

In this paper, we have established a connection between measure and
capacity for the space $\C$ of closed subsets of $\TN$. We showed that
for a computable measure $\mu^*$, a computable capacity may be defined
by letting $\T(Q)$ be the measure of the family of closed sets $K$
which have nonempty intersection with $Q$. We have proved an effective
version of the Choquet's theorem by showing that every computable
capacity may be obtained from a computable measure in this way.

We have established conditions on computable measures that
characterize when the capacity of a random closed set equals zero or
is $>0$. In particular, for symmetric measures where $d(\sigma \fr 0)
= d(\sigma \fr 1) = b \cdot d(\sigma)$ for all $\sigma$, where $b$
depends on $\sigma$, we have shown the following. If $d(\sigma \fr 2)
\leq \frac{\sqrt 2}2 d(\sigma)$ for all $\sigma$, then $\T_d(R) = 0$
for any $\mu_d^*$-random closed set $R$. If $d(\sigma \fr 2) \geq  
b \cdot d(\sigma)$ for all $\sigma$, where $b > \frac{\sqrt 2}2$ is a
constant, then $\T_d(R) > 0$ for any $\mu^*_d$-random closed set
$R$. 

We have shown that the set of capacities of an effectively closed set
is exactly the set of upper semi-computable reals. We have also
constructed effectively closed set with positive capacity and with
Lebesgue measure zero.

\end{document}